\documentclass[aps,prx,showpacs,amsmath,onecolumn,amssymb,superscriptaddress,letterpaper]{revtex4}
\usepackage{times}
\usepackage{amsfonts}
\usepackage{mathrsfs}
\usepackage{graphicx}
\usepackage{dcolumn}
\usepackage{bm}
\usepackage{color}

\usepackage[colorlinks,bookmarks=false,citecolor=blue,linkcolor=red,urlcolor=blue]{hyperref}
\usepackage{multirow}

\usepackage{color}
\usepackage{siunitx}
\usepackage{bibunits}

\def\be{\begin{equation}}       \def\ee{\end{equation}}
\def\bea{\begin{eqnarray}}      \def\eea{\end{eqnarray}}

\begin{document}


\title{Robustness of s-wave  pairing symmetry  in iron-based superconductors and its implications to fundamentals on magnetically-driven high temperature superconductivity}

\author{Jiangping Hu  }\email{jphu@iphy.ac.cn} \affiliation{ Institute of Physics, Chinese Academy of Sciences,
Beijing 100190, China}\affiliation{Department of Physics, Purdue University, West Lafayette, Indiana 47907, USA}
\affiliation{Collaborative Innovation Center of Quantum Matter, Beijing, China}
\author{Jing Yuan}\affiliation{ Institute of Physics, Chinese Academy of Sciences,
Beijing 100190, China}

\date{\today}

\begin{abstract}
Under the assumption that the superconducting state belongs to a single irreducible representation of lattice symmetry,  we argue that the pairing symmetry in all measured iron-based superconductors  is  universally  consistent with the $A_{1g}$ s-wave. The robust s-wave pairing throughout the different families of iron-based superconductors at different doping regions  signals two fundamental principles behind high $T_c$ superconducting mechanisms: (1) the  correspondence principle:  the short range magnetic exchange interactions and the Fermi surfaces act collaboratively to achieve high $T_c$ superconductivity and determine pairing symmetries; (2)  the magnetic selection pairing rule:  the superconductivity is only induced by the magnetic exchange couplings from the superexchange mechanism  through cation-anion-cation chemical bondings.   These principles explain why  the unconventional high $T_c$ superconductivity  appears to be such a rare but robust phenomena with  its strict requirement on  electronic environment: it can only be achieved on pure bands formed by the cation's d-orbital that have strongly in-plane chemical bonding with anions. The mixture of any other orbitals or destroying superexchange  rapidly suppresses superconductivity.    The robust s-wave pairing also  reveals  that  the  current standard effective models with only onsite interactions are not sufficient and  a minimum microscopic model must include strong nearest neighbor repulsive interactions resulted from the d-d direct bondings  to  serves as a s-wave symmetry stabilizer.  Finally,  the sign distribution of the superconducting order parameters in the reciprocal space is simply a consequence of the form factors given by   the leading short-range pairings. The  sign change of superconducting order parameters on Fermi surfaces is not a necessary  requirement in repulsive-interaction-driven high $T_c$ mechanism.  The results will guide us to search for new electronic structure that supports high $T_c$ superconductivity.
\end{abstract}

\maketitle
\section{Introduction}

The discovery of iron-based superconductors\cite{Kamihara2008-jacs} six years ago dethroned  the cuprates as the dictator of high $T_c$ superconductors in correlated electron systems and generated great hope and excitement to solve the decades-odd problem of non-BCS (Bardeen, Cooper and Schrieffer) high $T_c$  mechanism.  There were fundamental reasons for such an optimistic hope.  Iron-based superconductors were quickly revealed\cite{ironbook, Johnston2010-review,Dagotto2013-review} to share many common electronic properties with  the cuprates, including a close proximity to  an antiferromagnetic (AFM) order phase\cite{daihureview}. The similarity between the two superconductors strongly suggests  that there  should be one unified high $T_c$ superconductivity mechanism. In the meantime,  iron-based superconductors exhibit many distinct physical properties from the cuprates. The differences between these two materials  also provide us an opportunity to determine    side effects  that are irrelevant to  the superconducting mechanism.

However, the optimism was dying out in the past several years as more iron-based superconductors\cite{Johnston2010-review,Dagotto2013-review} were discovered and an unified understanding of all materials became increasingly difficult. In particular,   pairing symmetries  in iron-based superconductors have recently become more controversial than ever before\cite{Hirschfeld2011}. A variety of new possible pairing symmetries were proposed and many theories suggested that there is no universal pairing symmetry among iron-based superconductors\cite{Hirschfeld2011,Hu2013-odd,Hu2014-odd}. Namely,  pairing symmetries are  material and doping dependent.  This situation is very similar to the research status of the cuprates in the early 90's before the d-wave pairing symmetry was finalized\cite{Tsuei2000}. The d-wave pairing symmetry in the cuprates is widely acknowledged as  the major evidence to distinguish  the cuprates from conventional BCS-type s-wave superconductors. The robustness of the d-wave pairing  in  the cuprates is one of the main supports for magnetically-driven high $T_c$ superconducting mechanisms\cite{Scalapino1999,Anderson2004}. Thus, without a consensus on  the pairing symmetry in iron-based superconductors, it is difficult to imagine that the study of iron-based superconductors can help to advance the understanding of high $T_c$ superconducting mechanisms.

The pairing symmetry controversy in iron-based superconductors were largely caused by the ``top-to-bottom" theoretical studies. The study of  correlated  electron materials or materials that are believed to be in that category  has been customized to standardized models and methods\cite{Scalapino1999,Anderson2004}, inherited from the past  intensive research in cuprates.  After  iron-based superconductors were discovered, the extended versions of these standardized models, such as Hubbard-type\cite{Hirschfeld2011} or `t-J' -type models\cite{Si2008, Seo2008,sc-fang2011}, were immediately deployed.  The pairing symmetries in iron-based superconductors were analyzed with a variety of methods\cite{Hirschfeld2011}.  The theoretical predictions on the pairing symmetries are  in turmoil\cite{Hirschfeld2011}. In particular,  as Fermi surfaces  in iron-based superconductors can vary significantly in different doping regions, as well as in   different families of iron-based superconductors,   standard theoretical methods, due
to  their high sensitivity  to the change of Fermi surfaces,  suggest that  the pairing symmetries would vary significantly  as well.

However, if we take the ``bottom-up" approach and examine  the experimental evidence for the pairing symmetries in iron-based superconductors,   it is remarkable that  all experimental investigations that directly probe  the superconducting pairing symmetry point to an universal s-wave pairing symmetry in all measured materials\cite{arpes-review}, including the recent discovered materials\cite{FeSe-swave,FeSe-swave2,Zhao2015}.  Thus,  the robust s-wave pairing symmetry in iron-based superconductors is not only a challenge  but also  offers an opportunity to establish fundamentals in magnetically-driven high $T_c$ superconducting mechanisms.

\section{Pairing Symmetries and Gap functions}
Before we discuss the experimental evidence, for the sake of clarification, a few acceptable assumptions and notations should be  addressed.  First, we focus on  uniform superconducting state. Namely, the state has the original lattice translational symmetry. Second, we assume that the pairing  in iron-based superconductors is spin-singlet.  This assumption has been supported by many experiments throughout different families of iron-based superconductors\cite{Hirschfeld2011,Johnston2010-review,Dagotto2013-review}. Third,  we make clear that throughout the paper, unless it is otherwise specifically stated,  we use  the $1-Fe$ unit cell to label momentum space\cite{Leewen2008}.  In this case,  because the natural unit cell for iron-based superconductors is a $2-Fe$ unit cell,  in  the 11 ($FeSe$), 111 ($NaFeAs$) or 1111 $(LaOFeAs$) types of materials that have non-symmorphic space group $P4/nmm$ and the  122 ($BaFe_2As_2$) type of materials that has space group $ I4/mmm $,  the $\vec Q_1=(\pi,\pi,0)$ and   $\vec Q_2=(\pi,\pi,\pi)$ in the reciprocal lattice of the  $1-Fe$ unit cell are  reciprocal lattice vectors in the original $2-Fe$ unit cell\cite{Hu2013-odd}.  In the following part of paper, as we   primarily  focus on the two dimensional electronic structures of  the building block,  the $FeAs/Se$ layer, which is equivalent to setting $k_z=0$ for the bulk materials, both $\vec Q_1$ and $\vec Q_2$ are reduced to $\vec Q=(\pi,\pi)$.  Finally, as pointed out in Ref.\cite{Hu2013-odd},  there are two types of pairing that do not break translational symmetries with respect to the $2-Fe$ unit cell  in iron-based superconductors,  the normal pairing  which is between two particles with opposite momentums, namely, $(\vec k,- \vec k)$ and  the $\eta$ pairing which is specified by corresponding paired momentums, as $(\vec k,-\vec k+\vec Q)$. The difference between these two types of pairings are  classified by the parity difference with respect to the center on the nearest neighbour (NN) $Fe-Fe$ bonds.   Because of this difference, the mixture of normal and $\eta$ pairing  is also not a pure state.  Since a pure $\eta$ pairing is almost impossible under reasonable conditions, we will  ignore the $\eta$ pairing in the following  as well.

\begin{figure}[t]
\centerline{\includegraphics[height=8 cm]{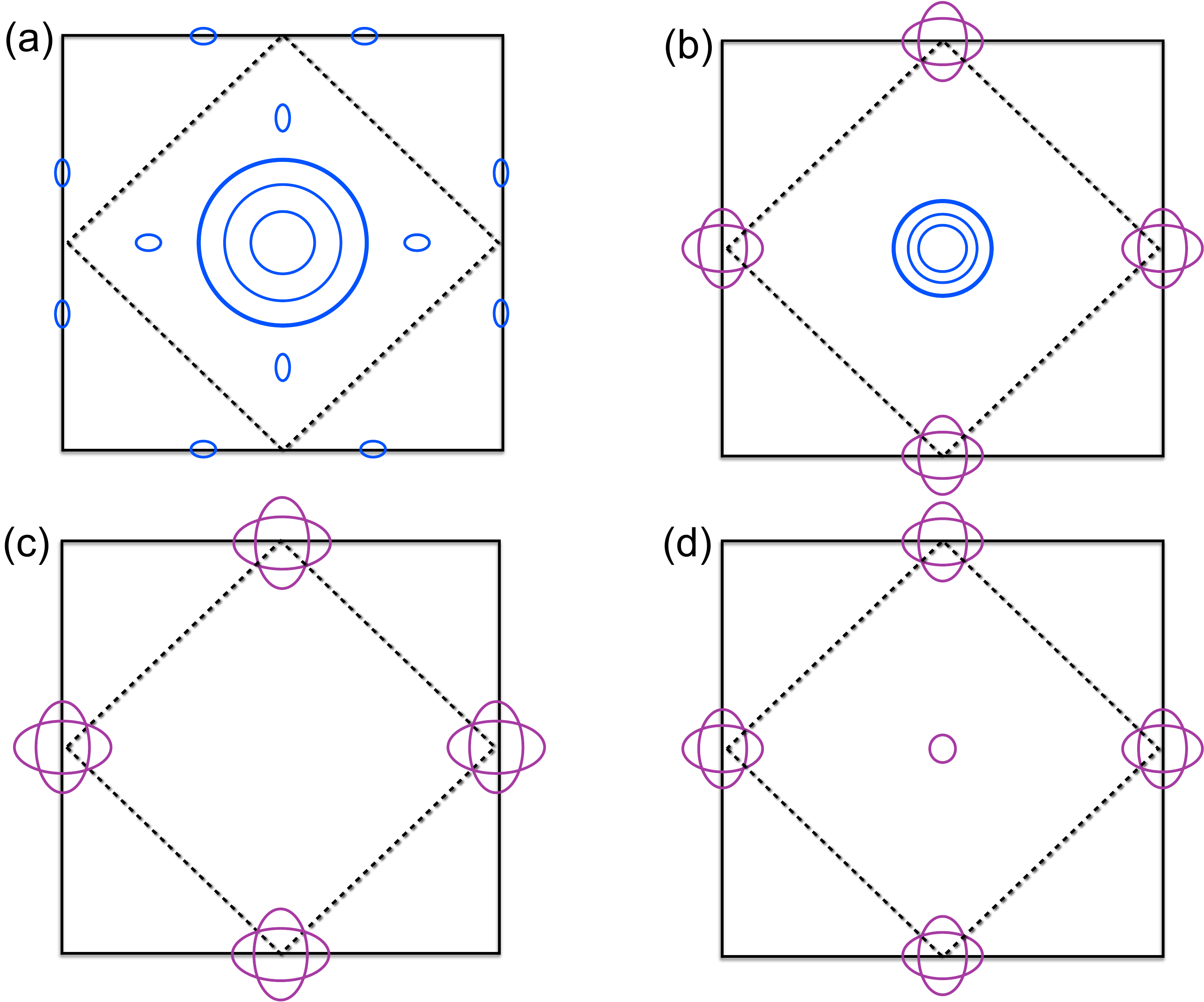}}
\caption{  The typical Fermi surface topologies of iron-based superconductors: (a) Heavily hole doped ($KFe_2As_2$); (b) Near optimally  doped iron-pnictides; (c) Heavily electron doped (single layer $FeSe/STO$, $(Li,Fe)OHFeSe$ and $KFe_2Se_2$ at $k_z=0$ ) ; (d) $KFe_2Se_2$ at $k_z=\pi$. }
\label{fig1}
\end{figure}

Assuming the superconducting state is a pure state, the pairing symmetry is manifestly reflected in superconducting gap functions in the reciprocal space. For iron-based superconductors,  if we take $k_z=0$ and ignore  spin-orbital couplings, which is known to be small\cite{Cvetkovic2013-band},  the normal pairing can be classified by the $D_{4h}$ group,  which is the same as cuprates.  Except the $A_{1g}$ s-wave, all other pairing symmetries have fixed gapless nodes along some high symmetry lines in the reciprocal space. In Fig.\ref{fig1},  we draw four typical Fermi surface topologies.  The Fig.\ref{fig1}(a)  represents  heavily hole doped iron-pnictides such as $KFe_2As_2$. There are three hole pockets at the zone center ($\Gamma$)  and four small hole pockets at the zone corner (M). With increasing electron doping,  the four hole pockets at the zone corner become two electron pockets as shown in Fig.\ref{fig1}(b) which represents the typical Fermi surfaces in iron-pnicitides within large doping regions.  It is also important to note that depending on the doping and materials, the number of hole pockets at the zone center as shown in Fig.\ref{fig1}(a,b)  can also vary. For example, in  the bulk $FeSe_xTe_{1-x}$ materials\cite{arpes-review}, there are only two hole pockets at the zone center and in the heavily electron doped $LiFeAs$\cite{arpes-review}, there could be just one hole pocket at the zone center.
With further increasing electron doping,  the hole pockets at the zone center can be suppressed completely as shown in the Fig.\ref{fig1} (c). The heavily electron doped iron-pnictides and  many iron-chalcogenides,  including the superconducting $KFe_2Se_2$\cite{Zhang2011}, the superconducting single layer $FeSe$ grown on the $SrTiO_3$ substrate\cite{Wang2012-fese,He2013-fese,Tan2013-fese} and  $(Li,Fe)OHFeSe$\cite{FeSe-swave,FeSe-swave2,Zhao2015},   have such a typical Fermi  surface topology. Away from  the $k_z=0$ plane,   heavily electron doped $KFe_2Se_2$ can also have a small three dimensional electron pocket as shown in Fig.\ref{fig1}(d) where a small electron pocket appears at the zone center at $k_z=\pi$\cite{Xu2012}.

In the presence of the hole  pockets at the zone center as shown in Fig.\ref{fig1}(a,b), a direct measurement of the gap functions on the hole pockets can determine the pairing symmetry. The angle resolved photoemission spectroscopy (ARPES) has measured a variety of iron-based superconductors\cite{arpes-review,Liu2015} that exhibit hole pockets at the zone center, including hole doped $Ba_{2-x}K_xFe_2As_2$, electron doped $Li(Na)Fe_{1-x}CoAs$,  isovalent doped $BaFe_2(As_{1-x}P_x)_2$ and
$BaFe_{2-x}Ru_xAs_2$, and $FeSe_{1-x}Te_{x}$. There are two common results on  the measured gap functions of  the hole pockets. \textit{ First,   the inner smallest hole pocket always has the largest full superconducting gap that is close to be isotropic. Second, the superconducting gap sizes on the hole pockets generally follow the rule: larger gaps on smaller hole pockets\cite{arpes-review}. } These two facts are  valid in all measured doping regions. If the superconducting state is a pure state, the two common results are only consistent with the s-wave pairing. In fact, even if we do not assume a pure state, these results suggest that the s-wave pairing must  be the dominant component.

In the case shown in Fig.\ref{fig1}(c,d),  in which there is no hole pockets at the zone center, a full gap structure on the electron pockets\cite{FeSe-swave,FeSe-swave2,Zhao2015,Zhang2011} was  also universally observed.  This full gap structure is generally only consistent with the s-wave pairing symmetry.  For a bulk material,  the nodes on electron pockets are inevitable if it is  a d-wave pairing state\cite{Mazin2011}. In the case of  Fig.\ref{fig1}(d),  which  represents  the Fermi surfaces of  $KFe_2Se_2$ at $k_z=\pi$,  there is another direct evidence for the s-wave pairing symmetry:  the small electron pocket  at  the zone center has almost isotropic full gap\cite{Xu2012,Zhang2011}.  A case that needs to be specifically addressed is the single layer $FeSe/STO$\cite{Wang2012-fese,He2013-fese,Tan2013-fese}.  In this case, a full gap structure on electron pockets, in principle, can  be consistent with the $B_{1g}$ d-wave. However, if we consider the hybridization between two electron pockets at the zone corner caused by the spin-orbital couplings and the lattice symmetry breaking  induced by the substrate,  gapless nodes must  generally appear in the $B_{1g}$ d-wave state.  Therefore, a full gap structure on electron pockets is also only consistent with a s-wave state.

Besides the direct gap function measurements of ARPES, the scanning tunneling microscopy(STM) also reveals full gap structure in many different materials\cite{Johnston2010-review}. The measurements on tunneling junctions also support the s-wave pairing symmetry\cite{ironbook}.

It is also important to mention that there is indirect evidence to support the d-wave pairing in some materials of iron-based superconductors. For example, possible d-wave pairing symmetry is indirectly indicated in $KFe_2As_2$ by thermal conductivity measurements and  pressure effect\cite{Tafti2015}. However, this indirect indication is not conclusive.   The thermal transport probes the existence of low energy excitations, namely,   the existence of superconducting nodes on Fermi surfaces. But it does not specify the structure of the nodes.  In fact, the ARPES measurements in $KFe_2As_2$  have shown that there are possible gapless nodes in the outer hole pockets\cite{Shin2014,Shin-kfese} but the inner hole pocket is fully gapped.  Nodes have also been directly observed by ARPES in $BaFe_2(As_{1-x}P_x)_2$ in the outer hole pocket at the zone center\cite{Zhang2012-node}.  Thus, the development of the accidental nodes appears to be tied with the enlargement of the outer hole pockets in  both  $KFe_2As_2$ and $BaFe_2(As_{1-x}P_x)_2$\cite{Lisy2012}, in which the pockets  reach close to the middle point of the first Brillouin zone.

In summary,  the full gap structure on Fermi pockets observed in a variety of iron-based superconductors at different doping regions is only consistent with the $A_{1g}$ s-wave pairing symmetry. The appearance of the largest isotropic gap on the smallest hole pocket suggests that the s-wave component dominates even if the superconducting state is not a pure state.

\section{ Theoretical results on pairing symmetries  of iron-based superconductors}
Through intensive research in cuprates, we have been accustomed to use two types of standard models to investigate correlated electron systems. The first type of standard models includes the band structure near Fermi surfaces and effective local repulsive interactions in the spirit of  the Hubbard model.   The second type of standard models includes an effective band structure and effective short-range magnetic exchange couplings in the spirit of the `t-J' model.  In cuprates, the d-wave pairing symmetry is  obtained consistently in both types of models\cite{Bickers1987,Inui1988,Cros1988,Kotliar1988,Metzner2012}.  In fact,  it is well-known that two models in cuprates are intimately linked.  There are also reasonable   microscopic derivations to support them in capturing essential physics of cuprates\cite{Zhang-rice}.

\begin{figure}[t]
\centerline{\includegraphics[height=8 cm]{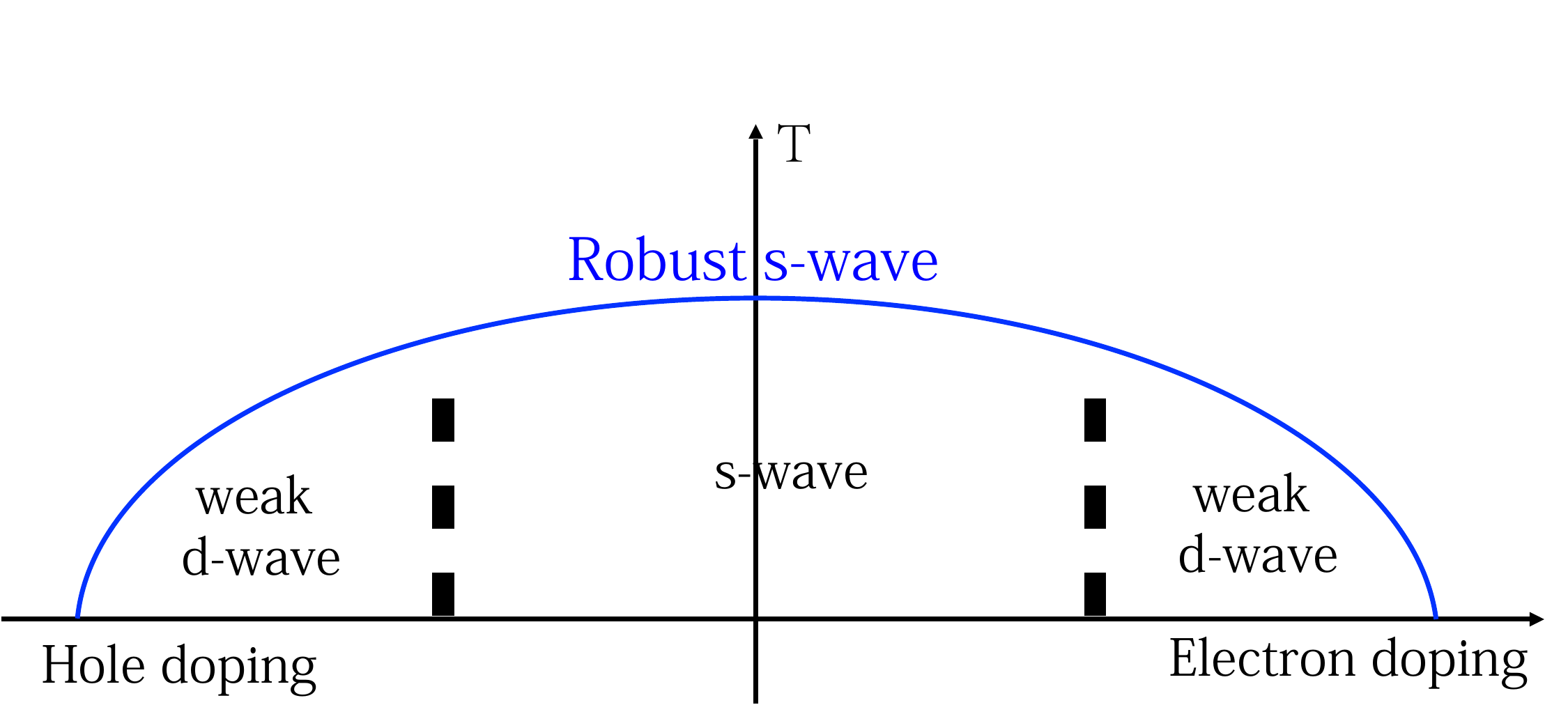}}
\caption{  The sketch of controversial pictures on pairing symmetries of iron-based superconductors as a function of doping: a robust s-wave pairing symmetry vs theoretical results as indicated by the black text) from the  $t-U-J_H$ model in Eq.\ref{tju}.   }
\label{fig2}
\end{figure}

However,  for iron-based superconductors the situation is rather different.  The first type of models in iron-based superconductors is generally proposed as\cite{Hirschfeld2011}
\begin{eqnarray}
 & \hat H_U &= \hat H_0+\hat H_{I},\\
 &\hat H_{I}&= \sum_{i} [\sum_{a\neq b}( J_H\hat S_{ai}\cdot \hat S_{bi}+U'\hat n_{ai\uparrow}\hat n_{bi\downarrow})+ U\hat n_{ai\uparrow}\hat n_{ai\downarrow}],
 \label{tju}
\end{eqnarray}
where $\hat H_0$ describes the effective multi-orbital band structures based on the five d orbitals of irons\cite{Graser2009-njp}, $a,b$ label orbitals and $i$ labels the sites of the iron square lattice. The  interactions include  all onsite interactions including the Hund's coupling $J_H$,    intra-orbital  repulsive  interaction $U$ and   inter-orbital repulsive interaction $U'$. Considering the interactions in the weak to intermediate regions, the model has been treated by a variety of approximate methods including  random phase approximation(RPA)\cite{Hirschfeld2011}, perturbative renormalization analysis\cite{Chubukov2011}  and numerical functional renormalization analysis (FRG)\cite{Wang2010}.  The pairing symmetries  obtained from the model  depend on the detailed structures of  Fermi surfaces. In general,  the strongest superconductivity is achieved  when  hole pockets and electron pockets are close to the nesting condition.  In this case, the pairing symmetry is  the  s-wave pairing symmetry, called $s_{\pm}$,  which is characterized by the sign change of the superconducting order parameters between the hole and electron pockets in momentum space.   With increasing hole (or electron) doping, the large anisotropic gap is expected to be developed on hole (or electron) pockets and a transition from the s-wave to d-wave pairing symmetry is  expected\cite{Maier2011,Thomale2011}.  In the Fig.\ref{fig2}, we sketch the phase diagram for the pairing symmetries from these  theoretical  studies.  Corresponding to  the Fermi surfaces shown in Fig.\ref{fig1},  the results suggest the d-wave pairing for Fig.\ref{fig1}(a,c,d)  and the s-wave  for Fig.\ref{fig1}(b).

A strong message from  the above theoretical studies  is that there is  no universal pairing symmetry in iron-based superconductors.  This conclusion is in direct conflict with the experimental observations. Moreover, within these studies, the strongest superconductivity is obtained  when  hole and electron pockets both exist and are close to the nesting conditions.  The theory clearly fails to explain  the existence of the high $T_c$ superconductivity  at heavily electron doped region of many $FeSe$-based compounds.

In the second type of models, the starting Hamiltonian\cite{Seo2008} is
\begin{eqnarray}
&\hat H_{tJ}&=\hat{\tilde{H}}_0+\hat H_{J}, \\
&\hat H_J&=\sum_{ij}J_{ij}\hat S_i\cdot \hat S_j,\\
\end{eqnarray}
where $\hat {\tilde H}_0$ describes the renormalized effective band structure and $\hat H_J$ describes the effective magnetic exchange couplings.    The pairing symmetry has been studied by including both NN magnetic exchange couplings, $J_{<ij>}=J_1$ and next nearest neighbour (NNN) ones, $J_{<<ij>>}=J_2$. Namely,  $H_J$ becomes
\begin{eqnarray}
\hat H_J= \sum_{<ij>}J_{1}\hat S_i\cdot \hat S_j+\sum_{<<ij>>}J_{2}\hat S_i\cdot \hat S_j.
\end{eqnarray}

Within this model, a robust s-wave pairing symmetry can be achieved if the NNN AFM exchange coupling $J_2$ dominates regardless of the presence or absence of the hole pockets\cite{sc-fang2011}. However, if the NN AFM exchange couplings are significant, the pairing symmetry is also expected to deviate from the s-wave\cite{Seo2008}.

The pairing symmetry in this model is simply determined by the \textit{`correspondence principle'} between Fermi surface topologies in reciprocal space and the symmetry form factors of short-range magnetic exchange couplings, which we refer  it  as \textit{ the HDLD  principle} since it has been  specified explicitly by  Hu-Ding\cite{Huding2012} and Lee-Davis\cite{Davis2014} recently. The principle can be applied to unify the understanding of pairing symmetries in both cuprates and iron-based superconductors.  In cuprates, the magnetic exchange coupling is dominated by the NN $J_1$. The d-wave form factor associated with $J_1$ in the pairing channel is given by $cosk_x-cosk_y$, which can open much larger superconducting gaps than the s-wave form factor $cosk_x+cosk_y$ on the Fermi surfaces of cuprates. In iron-based superconductors, the intra-orbital  pairing form factor associated with $J_2$ in the d-wave symmetry is $sink_xsink_y$, which has much smaller values on Fermi surfaces than the s-wave form factor $cosk_xcosk_y$. Thus, if $J_2$ is the dominating magnetic exchange coupling, the strong s-wave superconductivity can be achieved in both Fermi surface topologies shown in Fig.\ref{fig1}(b,c). However, it is clear that if $J_1$ is also important, the d-wave form factor $cosk_x-cosk_y$ associated with $J_1$ can also be very competitive when the electron pockets at the zone corner dominate. In Ref.\cite{sc-fang2011}, it is argued that  $J_1$ is inactive in spin-singlet pairing channel in iron-chalcogenides because $J_1$  based on neutron scattering experimental data is ferromagnetic (FM) so that the s-wave pairing prevails in all iron-chalcogenides\cite{Hu2012u,daihureview}.

In summary, the first type of the model, $\hat H_U$, does not support a robust s-wave. In the second type of model, $\hat H_{tJ}$, it  requires the dominance of $J_2$ over $J_1$ to establish a robust s-wave.  However,  the   assumption of the dominance in the second type of models is not well justified for  iron-pnictides where the $J_1$ value obtained from neutron scattering experiments can be strongly AFM.      Therefore,  the  robust s-wave symmetry in iron-based superconductors exposes serious limitation and deficiency  in current standard models based on  repulsive interaction (or magnetically) driven superconducting mechanisms.

\section{The  minimum microscopic models to stabilize s-wave pairing in  iron-based superconductors}
What are the deficiencies in the standard models for iron-based superconductors and how can they be fixed?  In the above section, we already notice that  the  robust s-wave can be obtained from the $J_2$  AFM exchange. This  provides an important clue. In fact, if we carefully check the electronic structure of iron-based superconductors, it is easy to see that  $J_1$ and $J_2$ involve different microscopic origins.

First,  the difference is already implied in  the low energy effective model that describes the magnetism in iron-based superconductors.   Extracted from neutron scattering experimental data, the value of  $J_2$ is quite universal throughout different families of  iron-based superconductors\cite{Hu2012u,daihureview}. However it is not the case for  $J_1$.   It has shown that the magnetism in iron-based superconductors can be unified  by the following  effective Hamiltonian,
\begin{eqnarray}
& & \hat H_{JK}= \sum_{<ij>}(J_{1}\hat S_i\cdot \hat S_j-K(\hat S_i\cdot \hat S_j)^2)\nonumber \\
& + &\sum_{<<ij>>}J_{2}\hat S_i\cdot \hat S_j+\sum_{<ij>TNN}J_{3}\hat S_i\cdot \hat S_j.
\end{eqnarray}
This model  was proposed as the minimum magnetic model to unify the magnetism in iron-based superconductors\cite{Hu2012u}.  Besides the $J_1$ and $J_2$ terms,  the model includes the quadruple spin interaction $K$ term between two  NN sites  and the third NN (TNN) magnetic exchange coupling $J_3$\cite{Ma2009fete}. The $K$ term was first proposed in Ref.\cite{Wysockio2011} to explain the large anisotropy  of the effective NN couplings along two different directions in the collinear AFM order state in iron-pnictides.

 There are several important observations related to $\hat H_{JK}:$ (a) the  classical phase diagram  of the model\cite{Hu2012u,Glasbrenner2015}  includes all observed magnetic long range orders, including both commensurate and incommensurate orders in different families of iron-based superconductors;
(b)    $J_2$ in all iron-based superconductors are universally AFM with a similar value while $J_1$ is not. In iron-pnictides, $J_1$ is AFM.  But in iron-chalcogendies, it changes to FM; (c) the model can even describe the observed magnetic orders in the  vacancy ordered states in $K_xFe_{2-y}Se_2$; (d) $J_3$ is significant  in iron-chalcogenides but small in iron-pnicitides; (e) the K term is significant  on the NN bonds.
The existence of the quadruple term  on the NN bonds and the non-universality of the $J_1$ value suggest that the mechanisms behind $J_1$ and $J_2$ are very different.

Second, there is a major difference between the electronic structures of iron-based superconductors  and those of cuprates. In cuprates,  without oxygen atoms,  the d-orbital of $Cu$ atoms  can be treated as  localized d-orbital (mootness).  The kinetic energy, namely, the effective hopping between  two d-orbital is caused by the d-p hybridization.  This hybridization simultaneously generates the magnetic superexchange couplings between two NN $Cu$ sites.  Therefore, the magnetism involves a pure superexchange mechanism.  If we consider an effective model based on the $Cu$ lattice,   the Hubbard term is sufficient to capture the superexchange coupling. Therefore,   magnetism and superconducting pairing symmetry can be consistently obtained in both `t-J'  and Hubbard  models.  However, in iron-based superconductors,  if we remove the $As/Se$ atoms in the $FeAs/Se$ layer and check the iron lattice, the distance between two NN irons is very short. In fact, the distance is very close to the lattice constant in  a three dimensional body centered iron metal.  Therefore, without $As/Se$ atoms, the large hopping between two NN $Fe$ atoms exists and the iron lattice itself  is a metallic state.  The chemical bonding  between two NN $Fe$ atoms can not be ignored.   The difference leads to several important consequences:  (1)  the hopping through the d-p hybridization that provides the superexchange mechanism is not the hopping defined in the effective Hubbard-type model given in Eq.\ref{tju}; (2) the NN hybridization between the d-orbital can also cause magnetism through direct exchange mechanism; (3) the overlap between two NN d-orbital also suggests that strong repulsive interactions between two NN sites have to  be included.   Thus, a model with only onsite repulsive interactions is not sufficient to capture the electronic physics of iron-based superconductors.   In an effective model with only onsite repulsive interactions given in Eq.\ref{tju},  $J_1$ and $J_2$ are both the leading  magnetic coupling terms and are developed  with equal footing.  Thus, the different magnetic origins between $J_1$ and $J_2$ are not taken into account by Eq.\ref{tju}.

The above analysis suggests that with only the simple onsite repulsive interaction, $\hat H_U$ in Eq.\ref{tju} is not sufficient to describe iron-based superconductors.  If we consider superconducting pairing driven by repulsive interactions,   the onsite repulsive interactions  only forbid onsite pairings. Without considering the NN repulsive interactions,    the NN pairings   generally  become the  leading contribution to pairing gap functions.   In  the model given by Eq.\ref{tju},  as both NN hoppings and NNN hoppings are large,
   $J_1$ and $J_2$  develop with  equal footing as  the leading  magnetic couplings to provide attractive pairing forces.  Because of their competition,  the pairing symmetry obtained from Eq.\ref{tju}  is highly sensitive to  the change of Fermi surfaces.

The above analysis also provide the solution to fix the problem.  The existence of chemical bonding between two NN $Fe$ d orbitals suggests that significant repulsive interactions between  the NN sites must exist in  the iron-based superconductors. This repulsive interaction can suppress the attractive interactions generated by the onsite interactions.  With both repulsive interactions on  onsite and between two NN  sites, the leading attractive forces must start  between the NNN sites so that the robust s-wave can be obtained.  Thus, a minimum term that can be added to    the Hubbard-type in Eq.\ref{tju}  is
\begin{eqnarray}
\hat H_{V}=\sum_{<ij>,ab}V_{ab}\hat n_{ia}\hat n_{jb},
\end{eqnarray}
 where $V_{ab}$ represents the repulsive interactions between two NN sites.

 Here, we are not interested in the detailed values of the NN repulsive interaction. Qualitatively,  we can argue that this repulsive interaction can serve a s-wave stabilizer.   The effect of this term on pairing can be understood in both real and momentum spaces. In the real space,  as repulsive interactions exist  both on onsite and between NN sites, the attractive force generated for pairing falls to the NNN  and  the TNN bonds. Namely, the dominated pairing in iron-lattice falls to pairing within each sublattice if we divide the iron lattice to two sublattice A and B.  The intersublattice pairing is suppressed. In the reciprocal space, the $\hat H_{V}$ allows  two Cooper pairs with pairing momentum $(\vec k,-\vec k)$ and $(\vec k+\vec Q,-\vec k+\vec Q)$  to be attractive.  Therefore,
the $s_{++}$ pairing can be robust in presence of only electron pockets at two M points.

\begin{figure}[t]
\centerline{\includegraphics[height=8 cm]{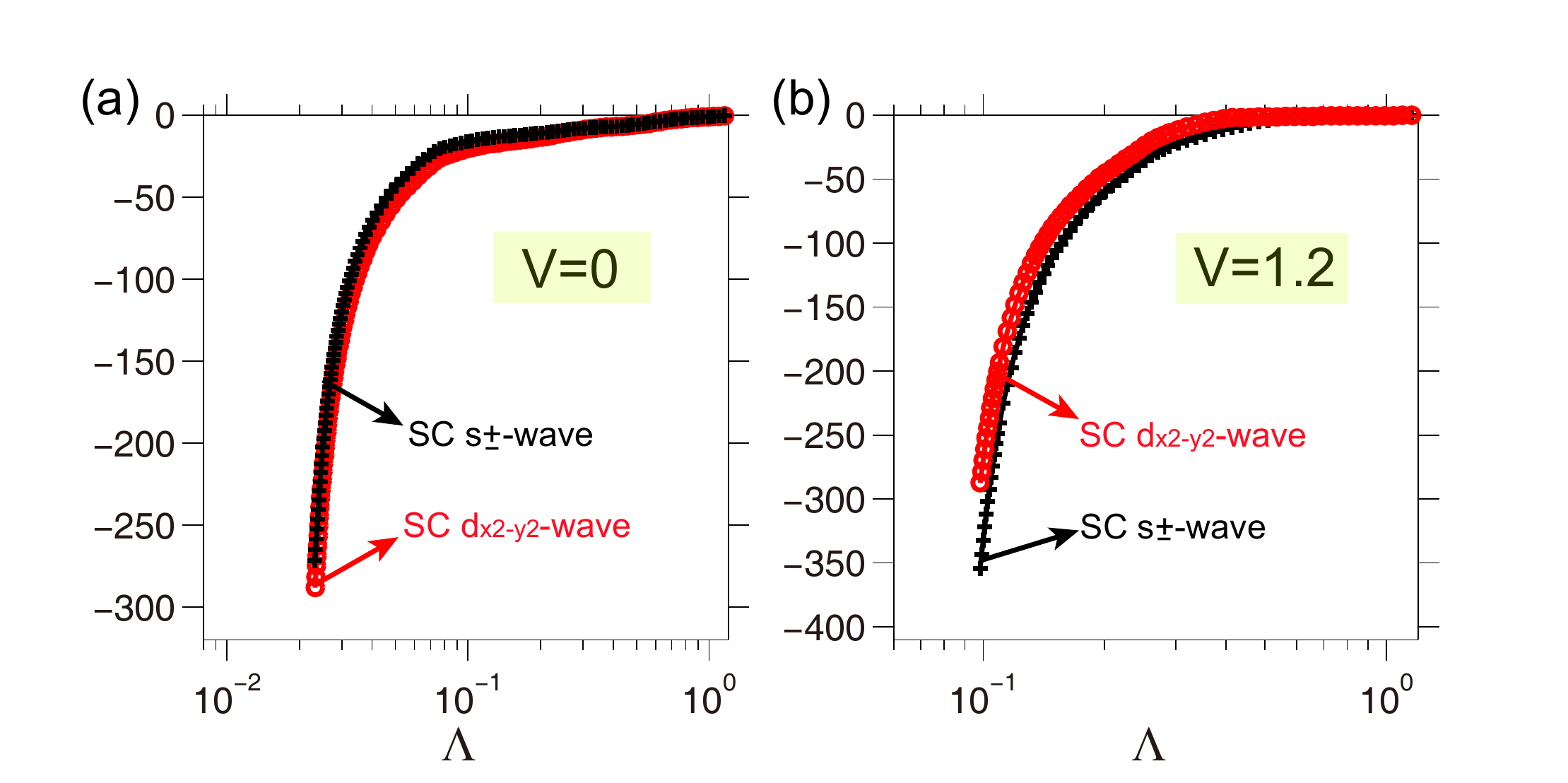}}
\caption{ The FRG flow of superconductivity s-wave and d-wave instabilities for five-band tight-binding model\cite{Graser2009-njp} with onsite repulsive interaction $U,U'$ and NN  repulsive interaction $V$: (a) $U=4,U'=4,V=0$, s-wave and d-wave symmetries are very competitive to each other. (b) $U=4,U'=4,V=1.2$, the divergence of s-wave is stronger  than d-wave after add $V$.  }
\label{fig3}
\end{figure}

To show that $\hat{H}_V$ can serve as a s-wave stabilizer even in a weak coupling approach, we performed a FRG calculation in a parameter region that s-wave and d-wave pairing symmetries are very competitive to each other. We take the band parameters provided by Ref.\cite{Graser2009-njp}. This five-band tight-binding model gives five Fermi surfaces when the chemical potential $\mu=-0.17$. In our calculation we set the interaction parameters as $U=4,U'=4$. The FRG flow demonstrates that s-wave and d-wave instabilities have approximately the same divergence as the momentum cutoff $\Lambda$ decreasing(see Fig.\ref{fig3}(a)). After adding the NN repulsive interaction $V$ into the interaction part, we obtain a significant enhancement of s-wave instability which indicates $\hat{H}_V$ tents to induce s-wave pairing symmetry. Fig.\ref{fig3}(b) shows the FRG flow when $V_{ab}=V=1.2$ for any $a$ and $b$, from it we can see that s-wave is distinctly stronger than d-wave.

\section{The emergence of the magnetic selection pairing rule and fundamentals on electronic structures for  high $T_c$ materials }
Unconventional high $T_c$ superconductivity appears to be a rare phenomena. Only two classes of materials were discovered accidentally by  intensive researches in the past several decades. However, once it happens, the superconductivity is very robust. These rareness and robustness have already spoken a fundamental difference between  unconventional high $T_c$ superconductors and  conventional BCS superconductors which  can almost ubiquitously take place in most metallic systems at low temperature.  The fundamental difference, as many of us believe today, is that the unconventional high $T_c$ superconductors belong to a new material category--correlated electron systems. Nevertheless, even if the materials are in a different category, explaining the rareness and robustness of high $T_c$ phenomena is still a fundamental problem as so many correlated electron systems have been discovered.
Theoretically, we have been accustomed to simplify  correlation physics into an onsite Hubbard interaction and have used it everywhere. If such a simplified model is applicable to all correlated  materials, it is difficult to understand  the rareness of high $T_c$ phenomena.

Combining the rareness  dillemma,  the robustness of pairing symmetries in both high $T_c$ superconductors,  the above analysis regarding the origin of the robust s-wave pairing symmetry in iron-based superconductors, we can only think of  one logistic answer, that is, \textit{high $T_c$ superconductivity  takes place in an electronic environment in which the paired electrons must  participate strongly in  magnetic superexchange mechanisms.   Namely,  only magnetic superexchange mechanism drives  superconducting pairing.}
With this hidden principle,  we can understand the rareness problem. Both the $d^9$ filling configuration in $Cu^{2+}$ in  cuprates and the $d^6$ filling configuration of $Fe^{2+}$ in iron-based superconductors are the unique configurations to make the orbital characters on the Fermi surfaces belong to the orbitals with strong involvement in superexchange processes.
Replacing $Cu$ in cuprates or $Fe$ in iron-based superconductors both violate this principle, which explain why the $Mn$, $Co$ and  $Ni$-based pnicitides are not high $T_c$ superconductors\cite{Ronning2008,Sefat2009,Singh2009}.  The principle also suggests that  the  effective standardized Hubbard and `t-J' models are only good approximations if all effective hopping parameters in the models  stem from the d-p hybridizations.  In fact, the presence of strong correlation effect  is generally correlated   together with  the existence of the superexchange processes.

  The above magnetic selection pairing rule, together with the HDDL principle, provides an powerful  guide to search for new unconventional high $T_c$ superconductors.  The rules are not easy to be satisfied in a three dimensional electronic structure.  Here, we focus on materials with a quasi-two dimensional layer structure constructed by transition metal cation-anion complexes. To be a high $T_c$ candidates,   the following  conditions must be followed:
  \begin{itemize}
\item \textit{The orbitals ( the d-orbital of cations ) responsible for high $T_c$  on Fermi surfaces must be strongly  coupled to  in-plane  anions.} This rule follows the fact that the superexchange is meditated through the  anions.The stronger coupling can produce stronger superexchange couplings, thus possible stronger superconductivity.  In  cuprates, the  $E_g$ orbital,  $d_{x^2-y^2}$, strongly couples to the p-orbital of in-plane oxygen. In iron-based superconductors, it is the three $t_{2g}$ orbitals which strongly couple to the p-orbital of $As/Se$.
\item \textit{The orbitals ( the d-orbital of cations ) responsible for high $T_c$  on Fermi surfaces should be relatively higher energy orbitals in local crystal field splitting environments.}  This rule follows that  the orbitals that  strongly couple to anions experience a larger crystal field energy.  This rule, thus, also implies that  cation atoms should  have high filling in their d-orbital shells in order to achieve possible high $T_c$. Namely, the second half transition metal cation atoms, which include both $Fe$ and $Cu$ atoms,   are more likely to  form potential high temperature superconductors. Of course, this rule does not completely rule out the possibility to achieve high $T_c$ for the first half transition metal cation atoms.
\item \textit{There should be no chemical bonding between  anions.} The chemical bonding between  anions can generally push the anti-bonding states to higher energy which can strongly suppress the superexchange processes.  This is the essential reason why the superconductivity and magnetism suddenly disappears in the collapsed $CaFe_2As_2$
phase where the chemical bondings between two $As$ atoms on neighbor layers are formed\cite{Yang2015}.
\item \textit{ The weight of other d-orbital on Fermi surfaces, which do not or only weakly couple to in-plane anions,  should be as small as possible.} The mixture of the other d-orbital can strongly suppress superconductivity. This fact has been known  in cuprates that raising the energy level of the  $d_{z^2}$ orbital  which increases  its' presence near Fermi surfaces strongly suppresses high $T_c$\cite{Sakakibara2013}.

\end{itemize}

In general, the above rules suggest that the electronic environment to host high $T_c$ superconductivity must include quasi-two dimensional bands  formed dominantly by the d-orbital through d-p couplings.

\begin{figure}[t]
\centerline{\includegraphics[height=6 cm]{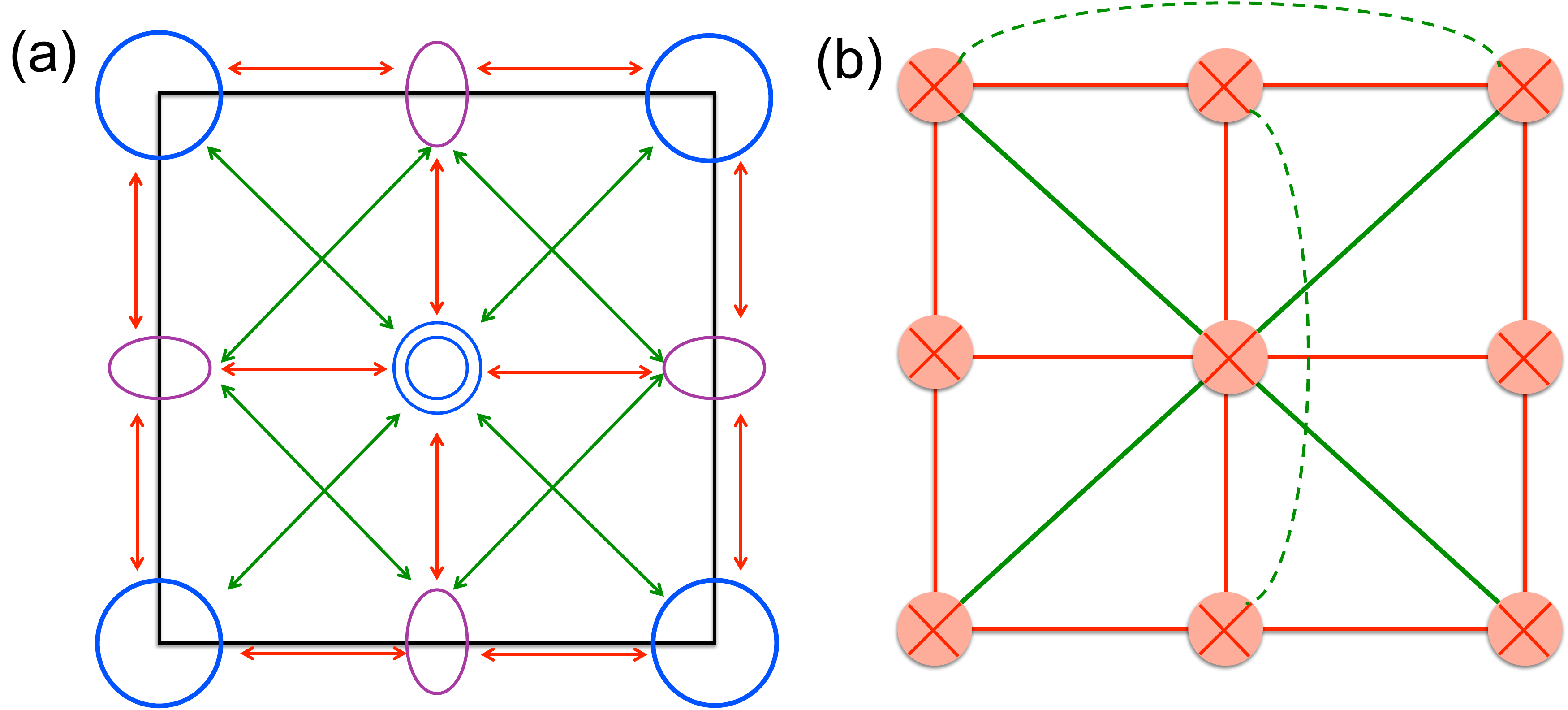}}
\caption{  (a) The pairing interactions between different pockets in the  1-$Fe$ unit cell Brillouin zone  for  iron-based superconductors ( the red and green arrows indicate repulsive and attractive pairing interactions respectively. (b)  The real space pairing configurations in the $Fe$ lattice for iron-based superconductors (only the two sites linked  by the green lines   are allowed to be paired.)   }
\label{fig4}
\end{figure}

 \section{Summary and Discussion}
In summary, the robustness of s-wave pairing in iron-based superconductors can only be understood by assuming that the pairing is exclusively formed through the superexhange process.  Namely, the superexchange AFM exchange couplings are responsible for superconducting pairing.  While this physics is very clear in cuprates,  it  is not obvious in iron-based superconductors because of the existence of the direct d-d bonding between two NN $Fe$ atoms.

Realizing the origin of pairing from exclusive superexchange AFM has an important impact on search new high $T_c$ materials and understand pairing properties. First, we should target to design or find a pure electronic lattice structure which host  bands with dominant contribution from d-orbital that involve  strong in-plane d-p hybridization. Second, the superconducting pairing can be better understood in real space than in momentum space as the superexchange process is a local process. Fig.\ref{fig4}(a,b) sketches the leading pairing bonds for iron-based superconductors  in real space and its corresponding pairing interaction picture in momentum space.  A natural consequence of  the real space picture is that   the sign change  of the superconducting order parameters on Fermi surfaces is not a necessary requirement.  This implication also suggests that the earlier argument in the weak coupling approach about the $s_{\pm}$ pairing symmetry that is based on momentum space  is not completely correct\cite{Mazin2008}. Third,   it is easy to understand why the superconductivity is so robust but in the meanwhile is so sensitive to the lattice parameter change of the anions as the anions  essentially mediates superconducting pairing.   For example, the $Fe-As-Fe$ angle is a critical parameter in determining $T_c$ in iron based superconductors\cite{Johnston2010-review}. Finally, the principles also provide an intuitive explanation about why high $T_c$ superconductivity is absent in many materials with strong magnetic fluctuations.

\textbf{Acknowledgement: } { The work is supported by  the National Basic Research Program of China, National Natural Science Foundation of China (NSFC)
and the Strategic Priority Research Program of  Chinese Academy of Sciences. }

\bibliography{hightc}

\end{document}